# Unlocking Altermagnetism in Antiferromagnetic 2D Films via Adsorption

Dong Liu, Sike Zeng, Ji-Hai Liao and Yu-Jun Zhao*

Department of Physics, South China University of Technology, Guangzhou 510640, China

*Corresponding author: zhaoyj@scut.edu.cn

**ABSTRACT**: Altermagnets, characterized by zero net magnetization and momentum-dependent spin splitting, have recently garnered significant attention due to their potential applications in a variety of fields. Here, we propose a symmetry-engineering strategy to unlock altermagnetism in two dimensional (2D) antiferromagnetic systems via surface adsorption of atoms or molecules. By employing spin group theory, we systematically demonstrate that selectively breaking symmetry operations, specifically those protecting spin degeneracy in momentum space, enables the emergence of nonrelativistic spin-split electronic states. Meanwhile, preserving rotation or mirror symmetries connecting opposite sublattices ensures zero net magnetization. Through a comprehensive classification of all symmetry operations across 80 layer groups, we identify 63 antiferromagnetic spin point groups (SPGs) describing 2D materials and further isolate 15 groups that can host altermagnetic characteristics through surface adsorption. Exemplified with monolayer antiferromagnetic $VPS_3$ and $MnPSe_3$, we show that oxygen adsorption on $VPS_3$ and $NH_3$ adsorption on $MnPSe_3$ selectively disrupt $PT$ symmetry while retaining the $[C_2||m]$ symmetry. This engineered symmetry reduction induces pronounced spin splitting in their band structures without spin-orbit coupling, as confirmed by first-principles calculations. Furthermore, adsorption energy analysis and thermal stability phase diagrams under varying coverage regimes reveal optimal configurations for experimental feasibility. Our work establishes a universal symmetry-engineering framework to expand the family of altermagnetic materials, offering a versatile pathway to tailor spin-split functionalities in two-dimensional antiferromagnets for advanced quantum applications.

## I. INTRODUCTION

Altermagnetism, a distinct class of antiferromagnetism, is characterized by momentum-space alternating band splitting and an overall vanishing net magnetic moment [1,2]. It has garnered significant recent attention due to its fundamental physics and potential device applications [3-10]. In altermagnets, antiparallel spin sublattices are connected not by translation or inversion, but by proper or improper rotations. This symmetry leads to nonrelativistic band splitting and broken time-reversal symmetry, while the symmetry connecting opposite-spin sublattices ensures momentum-space alternating spin polarization and a net zero magnetic moment. Crucially, the band splitting in altermagnets is of nonrelativistic origin, implying that the splitting magnitude can be substantial even in the limit of negligible spin-orbit coupling (SOC), as evidenced in materials like CrSb [5], MnTe [11,12], and $MnF_2$ [13]. This unique phenomenology enables intriguing effects such as the anomalous Hall effect [14], magneto-optical Kerr effect [15] and giant magnetoresistance (GMR) [16]. For applications, altermagnets hold considerable promise in spintronics due to advantages including the absence of stray fields, terahertz spin dynamics, and strong time-reversal symmetry ($T$) breaking.

Due to the advantages of an atomically thin structure and high tunability inherent in two-dimensional materials, two-dimensional (2D) altermagnets have recently garnered increasing attention [6]. Expanding the family of 2D altermagnets would significantly enhance our fundamental understanding of their properties and accelerate their application in spintronics. Consequently, considerable effort is directed towards enriching the candidate pool for 2D altermagnetism. Previous studies employed methods such as applying vertical electric fields [17], strain [18], constructing Janus structures [8], or specific stacking configurations to realize A-type altermagnetism [19]. These approaches modify the symmetry connecting antiparallel spin sublattices to satisfy the specific symmetry requirements of altermagnetism, thereby broadening the range of potential 2D altermagnetic materials. However, there are still some issues that remain unresolved. Most studies to date remain theoretical, and experimental investigations are still scarce. As a common experimental technique, adsorption can induce altermagnetism in 2D antiferromagnets, which remain unexplored. Moreover, a general theoretical framework for achieving 2D altermagnetism from antiferromagnetic systems remains lacking.

In this work, we enumerate all possible spin point groups (SPGs) of 2D antiferromagnetic systems. We construct 63 SPGs for antiferromagnets and 15 distinct SPGs capable of realizing altermagnetism through surface adsorption. Moreover, we demonstrate adsorption as an effective symmetry-breaking mechanism to induce the altermagnetic state in two 2D antiferromagnets, using $MnPSe_3$ and $VPS_3$ as representative examples. Further stability analysis demonstrates its experimental feasibility.

## II. DERIVATION OF 2D ANTIFERROMAGNETIC SPGs

To systematically characterize the symmetry breaking induced by adsorption and to identify potential candidate materials, we develop a theoretical framework that achieving 2D altermagnetism from conventional antiferromagnets through symmetry breaking. As a first step, we introduce the spin point groups, which have been proposed in the previous studies [20]. Nevertheless, when applied to (quasi-)two-dimensional systems, several critical differences arise. For example, the SPG $^22/^2m_z$ describes antiferromagnet, while the SPG $^22/^2m_x$ describes altermagnet. However, the SPGs in previous studies only have $^22/^2m$. While point group operations are inherently non-directional, such a treatment is more suitable in the present context. Accordingly, we begin by identifying the spin point groups that describe 2D conventional antiferromagnets and 2D altermagnets. Based on this, we analyze the symmetry breaking induced by surface adsorption.

A spin group can be expressed as the direct product $r_s \otimes R_s$. Here, the spin-only group $r_s$ acts exclusively on the spin degrees of freedom, while the nontrivial spin group $R_s$ comprises paired operations $[R_i||R_j]$, in which the transformation to the left of the double bar acts solely in spin space and the transformation to the right acts in real space. For collinear spin orderings, the spin-only group is identical and consists of only two symmetries: (i) all rotations in spin space about the common spin quantization axis $C_\infty$; and (ii) a twofold rotation about an axis perpendicular to the spin direction, combined with a spin space inversion—which is always accompanied by time-reversal symmetry $T$ and can be written as $[\bar{C}_2||T]$.

We focus on the nontrivial spin groups in the discussion below, because whether an antiferromagnet is altermagnet or not is only dependent on them [1,21]. For collinear spin configurations, the relevant spin-space operations of the nontrivial spin groups are chosen as the identity $E$ and a twofold rotation $C_2$ about an axis perpendicular to the spin direction. Within the framework of spin point groups, the $TT$ antiferromagnet is described by the spin group $R = [E||G] + [C_2||G]$; $PT$ antiferromagnet by $R = [E||H] + [C_2||P][E||H]$, where $P$ is the space inversion symmetry. Type-IV two-dimensional collinear magnet [22] is described by $R = [E||H] + [C_2||A][E||H]$, in which A is not space inversion but the twofold rotation around the z axis $C_{2z}$ or the mirror symmetry through the xy plane $m_z$. The spin degeneracy in the band structures of these antiferromagnets arises from their underlying symmetry properties. In addition to these antiferromagnets, there exist antiferromagnets with spin-split band structures, known as altermagnets, characterized by a spin point group $R = [E||H] + [C_2||A][E||H]$, where A represents a proper or improper rotational operation excluding P, $C_{2z}$ and $m_z$.

This work focuses on realizing altermagnetism via surface adsorption specifically within the framework of the spin-degenerate antiferromagnetic phase. Our classification builds on previous studies of SPGs [20]. To construct SPGs for 2D antiferromagnets, we first extracted point group operations from the 80 layer groups [23,24]. From these, we derived 63 antiferromagnetic SPGs, as listed in Table S1 and S2. Among these, 26 SPGs are identified as altermagnets, leaving 37 SPGs hosting spin degeneracy. By further determining whether surface adsorption can break the symmetries protecting spin degeneracy while preserving rotation or mirror symmetries connecting opposite-spin sublattices, we identified 15 SPGs where altermagnetism can be induced through adsorption. For instance, the spin point group $^22/^1m_x$ is constructed from layer group 14-18. To achieve altermagnetism, the symmetry $[C_2||\bar{E}]$, which protects spin degeneracy by connecting opposite-spin sublattices, must be broken while the symmetry $[C_2||C_{2x}]$ is preserved. However, surface adsorption simultaneously breaks both $[C_2||\bar{E}]$ and $[C_2||C_{2x}]$. Thus, such antiferromagnets cannot exhibit altermagnetism via surface adsorption. The remaining 15 SPGs, corresponding to antiferromagnetism, permit altermagnetic realization via surface adsorption, as listed in Table 1, with the adsorption sites enabling altermagnetism provided in the third column.

To facilitate the search for candidate antiferromagnetic materials amenable to surface adsorption induced altermagnetism, we summarize the material discovery and realization process into a comprehensive flowchart, presented in Fig.1. Furthermore, we later demonstrate how this flowchat guides the realization of altermagnetism in specific antiferromagnetic materials.

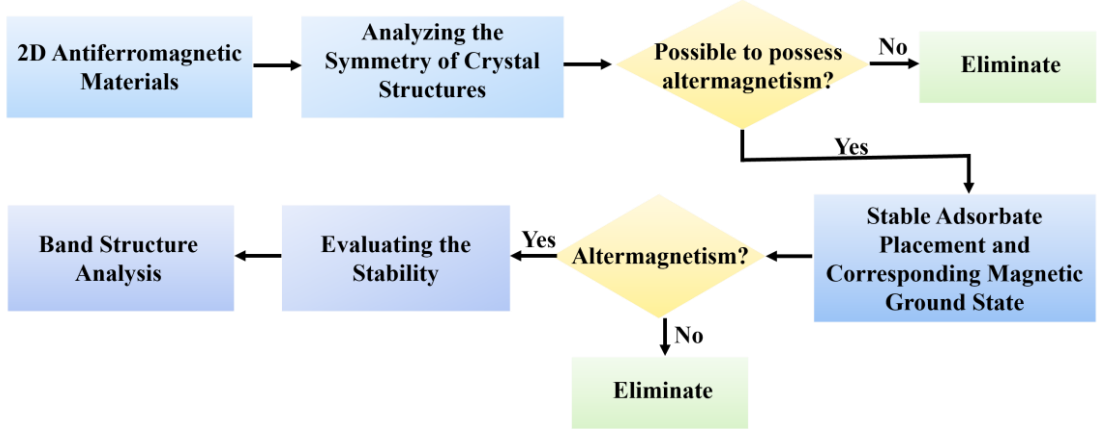

Fig. 1 Workflow for identifying antiferromagnets in which surface adsorption induces alter magnetism and for implementing the proposed adsorption strategy.

Table I Layer groups enabling altermagnetism via surface adsorption, their corresponding spin point groups (using Litvin's notation), and the requisite adsorption site symmetries (Column 3). A bar above a symbol denotes inversion symmetry. For example, $\bar{3}$ denotes 120° rotation with space inversion. The symmetry elements m and $C_{nz}$ denote a mirror plane and an $n$-fold rotation axis perpendicular to the $xy$-plane (assumed parallel to the material surface), respectively. Adsorption sites labeled $C_{nz}$ require the adsorbate to reside precisely on the corresponding $n$-fold $z$-axis rotation symmetry element.

| Layer groups | Spin point groups | Adsorption site |
|---|---|---|
| 14-18 | $^1 2/^2 m_x$ | $m_x$ |
| 23-26 | $^1 m^2 m^2 2$ | $m$, not $C_{2z}$ |
| 27-36 | $^2 m^1 2^2 m$ | $m_x$ |
| 37-48 | $^1 m^2 m^2 m$ | $m$, not $C_{2z}$ |
| 37-48 | $^1 m^2 m^1 m$ | $m$, not $C_{2z}$ |
| 37-48 | $^2 m^2 m^2 m$ | $m$ |
| 51-52 | $^2 4/^2 m$ | $C_{4z}$ |
| 61-64 | $^2 4/^2 m^2 m^1 m$ | $C_{4z}$ |
| 61-64 | $^1 4/^2 m^2 m^2 m$ | $C_{4z}$ |
| 71-72 | $^2 \bar{3}/^2 m$ | $C_{3z}$ or $m$ |
| 77 | $^2 6^1 m^2 m$ | $m$, not $C_{2z}$ |

| 78 | $^2\bar{6}^2m^12$ | $C_{3z}$ |
| 80 | $^26/^2m^1m^2m$ | $m$ |
| 80 | $^26/^1m^2m^1m$ | $m$ |
| 80 | $^16/^2m^2m^2m$ | $C_{6z}$ |

### III. MATERIAL CANDIDATES

Using monolayer $VPS_3$ and monolayer $MnPSe_3$ (belonging to layer group 72) as examples, both exhibit a Néel-type magnetic ground state [25,26], resulting in SPG $^2\bar{3}/^2m$. Here, the symmetry operations connecting the opposite-spin sublattices are $[C_2||\bar{E}]$, $[C_2||\overline{C_{3z}^+}]$, $[C_2||\overline{C_{3z}^-}]$, $[C_2||m_{1\bar{1}0}]$, $[C_2||m_{120}]$ and $[C_2||m_{210}]$, as illustrated in Fig. 2. To realize altermagnetism, $[C_2||\bar{E}]$ must be broken, while at least one of the remaining symmetries must be preserved. As indicated in Table I, altermagnetism can be realized in these materials via adsorption on mirror planes or rotation axes, as illustrated in Fig. 2(c). Below, we demonstrate how surface adsorption induces alternating spin splitting from their spin-degenerate bands.

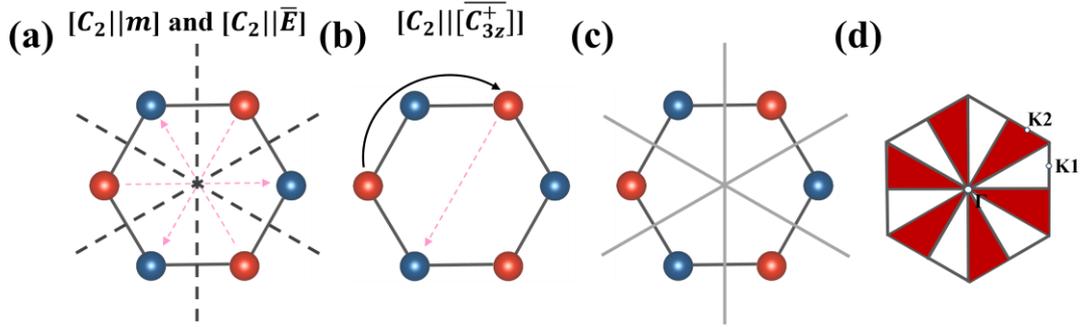

Fig. 2 Symmetry operations connecting opposite-spin sublattices in monolayer $VPS_3$ and $MnPSe_3$: (a) $[C_2||m_{1\bar{1}0}]$, $[C_2||m_{120}]$, $[C_2||m_{210}]$, $[C_2||\bar{E}]$ and (b) $[C_2||\overline{C_{3z}^+}]$. (c) Adsorption position (gray solid line) capable of inducing altermagnetism. (d) Resulting alternating spin polarization in momentum space after adsorption. Red and blue spheres denote spin-up and spin-down, respectively. A sequence of three numbers denotes the direction of the axis or plane.

The atomic structure of monolayer $VPS_3$ is presented in Fig. 3(a). The optimized lattice constants of $a = b = 5.99$ Å for monolayer $VPS_3$ are close to the experimental bulk values of 5.85

Å [27]. The calculated magnetic ground state is Néel-antiferromagnetic (Néel-AFM), with an energy difference $\Delta E_{AFM-FM}$= -0.16 meV/formula unit (f.u.), in agreement with experimental observations [25]. As shown in Fig. 3(b), the bands of pristine $VPS_3$ exhibit complete spin degeneracy throughout the Brillouin zone. Its SPG is $^2\bar{3}/^2m$, suggesting that it is possible to possess altermagnetism via adsorption onto mirror planes or symmetry axes, according to Table I. Specifically, the spatial inversion symmetry is broken, while the mirror symmetry linking opposite sublattices remains intact. This approach can consequently induce alternating spin polarization in momentum space.

Following this strategy, we considered three distinct adsorption sites (atop V, P, and S atoms), as illustrated in Fig. 3(a). Calculations employed four supercell sizes corresponding to coverages of 25%, 50%, 75%, and 100% (Fig. S1). Structural optimization identifies the P-atop site as energetically most favorable for oxygen adsorption for all coverages. Crucially, this adsorption preserves the Néel − AFM ground state. Figure S2 presents the band structures of monolayer $VPS_3$ after O adsorption at all coverages. Notably, the 100 % coverage case exhibits the largest band-splitting magnitude, corresponding to a Bader charge transfer of 0.63 e/formula unit (f.u.). In contrast, no clear correlation emerges between splitting magnitude and charge transfer at lower coverages.

Fig. 3(c) and (d) displays the structural configurations and electronic band structures after oxygen adsorption at 100% coverage. The spin point group of the $VPS_3$ monolayer transforms from $^2\bar{3}^2m$ to $^13^2m$, corresponding to the altermagnetic SPG. Compared to Fig. 3(b), Fig. 3(d) clearly demonstrates that oxygen adsorption lifts the original spin degeneracy, inducing pronounced alternating spin-splitting in the band structure.

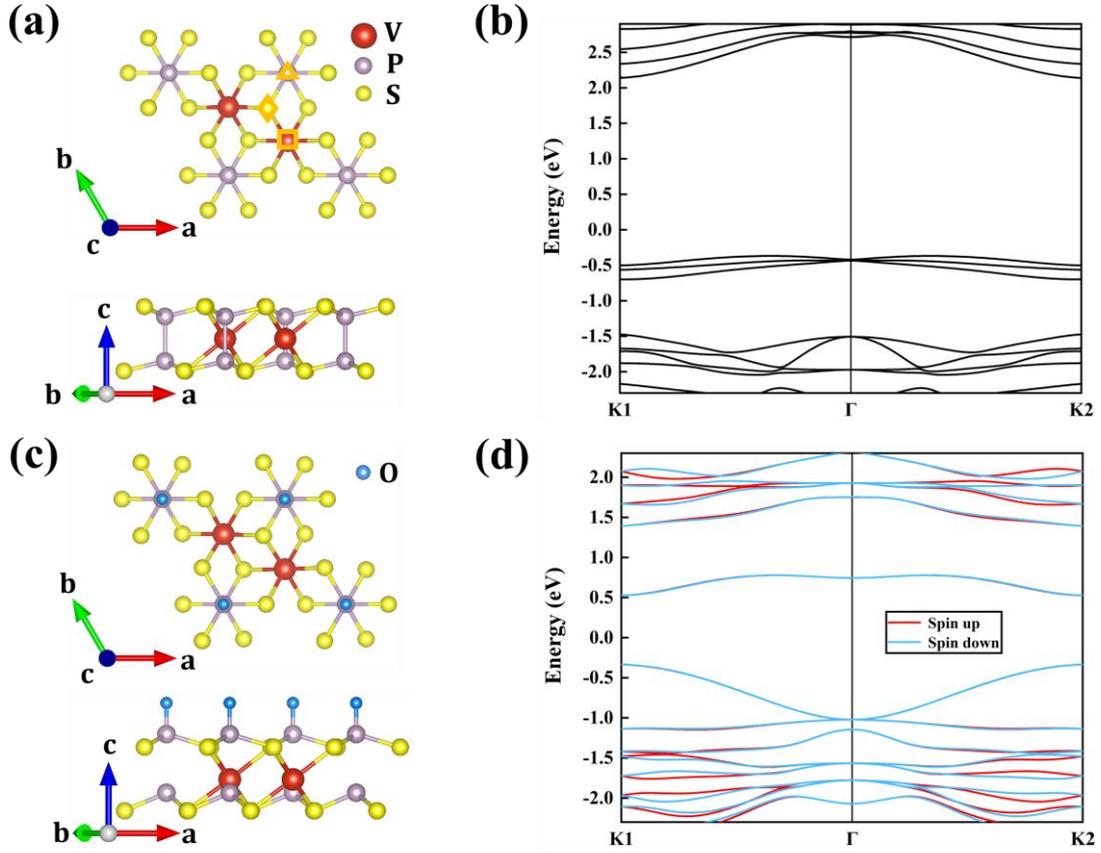

Fig. 3 (a) Top and side views of the initial monolayer $VPS_3$ structure. (b) Corresponding band structure without SOC. (c) Most energetically favorable configuration for O atom adsorption on the monolayer $VPS_3$ surface at 100% coverage. (d) Corresponding spin-splitting band structure without SOC. The square, triangle, and rhombus symbols denote the three considered adsorption sites for V, P and S. The K path used to calculate the band structure in (b) and (d) is shown in Fig. 2(d).

To verify the feasibility of adsorption, we calculated the adsorption energy for each atom using the following formula:

$$E_{ad} = E_{total} - E_{VPS_3} - E_i. \qquad (1)$$

Here, $E_{VPS_3}$ denotes the total energy of the pristine monolayer, $E_{total}$ corresponds to the energy of the adsorbate-substrate system, and $E_i$ represents the energy of an isolated *i* atom in its free state.

Table S3 presents the Bader charge transfer and corresponding adsorption energies for O at various coverages, alongside the cohesive energy of $O_2$ ($E_{coh}$ (O) = −3.04 eV), confirming that O adsorption on monolayer $VPS_3$ does not favor bulk-oxide formation. Based on the adsorption energy data listed in Table S3, oxygen adsorption at 100% coverage is found to be the most favorable.

Phase diagram of the V-P-S system under equilibrium conditions were computed by evaluating competing binary phases, yielding permissible chemical potential ranges (blue region in Fig. 4a). Furthermore, thermodynamic stability across all coverages was assessed by comparing the chemical potential of adsorbed oxygen with those of competing binary phases (see Supporting Information for details). Calculations reveal that all configurations lie within the thermodynamically stable region, as shown in Fig. 4(b).

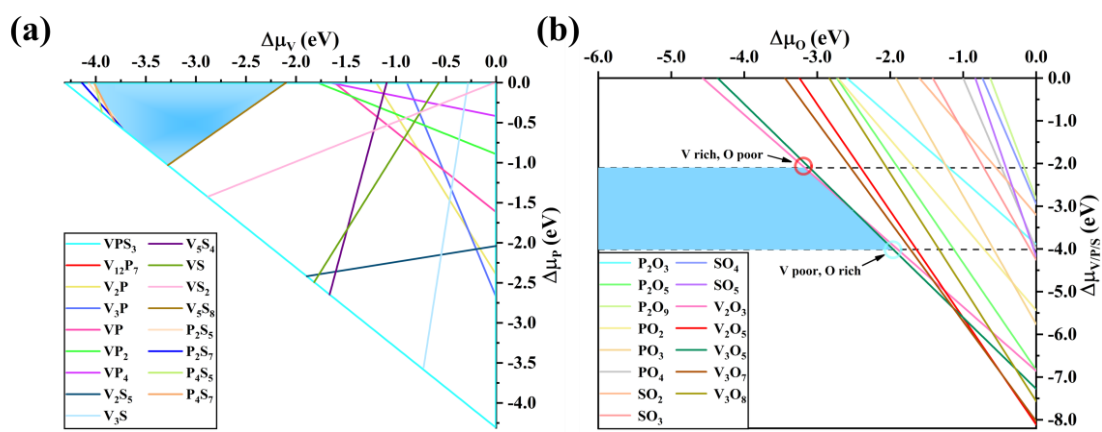

Fig. 4 (a) Parametrized phase diagram of $VPS_3$ projected onto the ($\Delta\mu_V$, $\Delta\mu_P$) plane. The blue region denotes the thermodynamically permissible range of V, P, and S chemical potentials under equilibrium conditions; (b) Chemical potential stability region for O adsorption. The blue region defines stable O chemical potentials ($\Delta\mu_O$).

The specific structure of the monolayer $MnPSe_3$ is depicted in Fig. 5(a). As shown, the two-dimensional $MnPSe_3$ layer comprises a hexagonal honeycomb lattice of $Mn^{2+}$ ions, coupled with a $[P_2Se_6]^{4-}$ bimetallic pyramid. This structural unit arises from a P–P dimer coordinated to two $Se_3$ trimers, positioned vertically through the center of each honeycomb plane. The crystal belongs to the $P\bar{3}m1$ space group. Our calculated lattice constants for monolayer $MnPSe_3$ are $a = b = 6.46$ Å, in agreement with the recently reported value of $a = b = 6.39$ Å [26]. The magnetic ground state is Néel-AFM, with an energy difference $\Delta E_{AFM-FM} = -21.9$ meV/formula unit (f.u.). As derived in Table I, its SPG is $^2\bar{3}/^2m$, suggesting that altermagnetism could be induced by breaking spatial inversion symmetry, potentially via adsorption onto mirror planes or symmetry axes.

Notably, we find that adsorption of atoms such as Li, O, and F on the monolayer $MnPSe_3$

surface induces an AFM-to-FM magnetic phase transition [28]. This result aligns with previous studies demonstrating magnetic phase transitions in $MnPSe_3$ driven by charge carrier doping [29]. To this end, we explore molecular adsorption as a route to break *PT* symmetry. Three distinct adsorption sites were considered: atop Mn, P, and Se atoms, as illustrated in Fig. 5(a). Calculations employing two different supercell sizes (1×1×1 and 2×2×1), corresponding to different $NH_3$ coverage densities, consistently identified the site atop the P atom as energetically most favorable for all coverages. At this site, the three N–H bonds are oriented precisely along the mirror planes. Initial perturbations applied to displace the N–H bonds away from these mirror planes invariably relaxed back to the symmetric orientation during optimization.

The calculated adsorption energies for $NH_3$ on $MnPSe_3$ are -0.12 eV and -0.16 eV per molecule for coverages of 25% and 100%, respectively, indicating enhanced stability at higher coverage. For clarity, Fig. 5 presents the structural configurations and electronic band structures before and after adsorption at 100% coverage. The corresponding data for the 25% coverage case are shown in Fig. S3. Crucially, Fig. 5(d) reveals a clear lifting of the spin degeneracy in the electronic bands following $NH_3$ adsorption. This spin-splitting, exhibiting an alternating pattern between bands, is a direct consequence of the broken spatial inversion symmetry induced by the adsorbate.

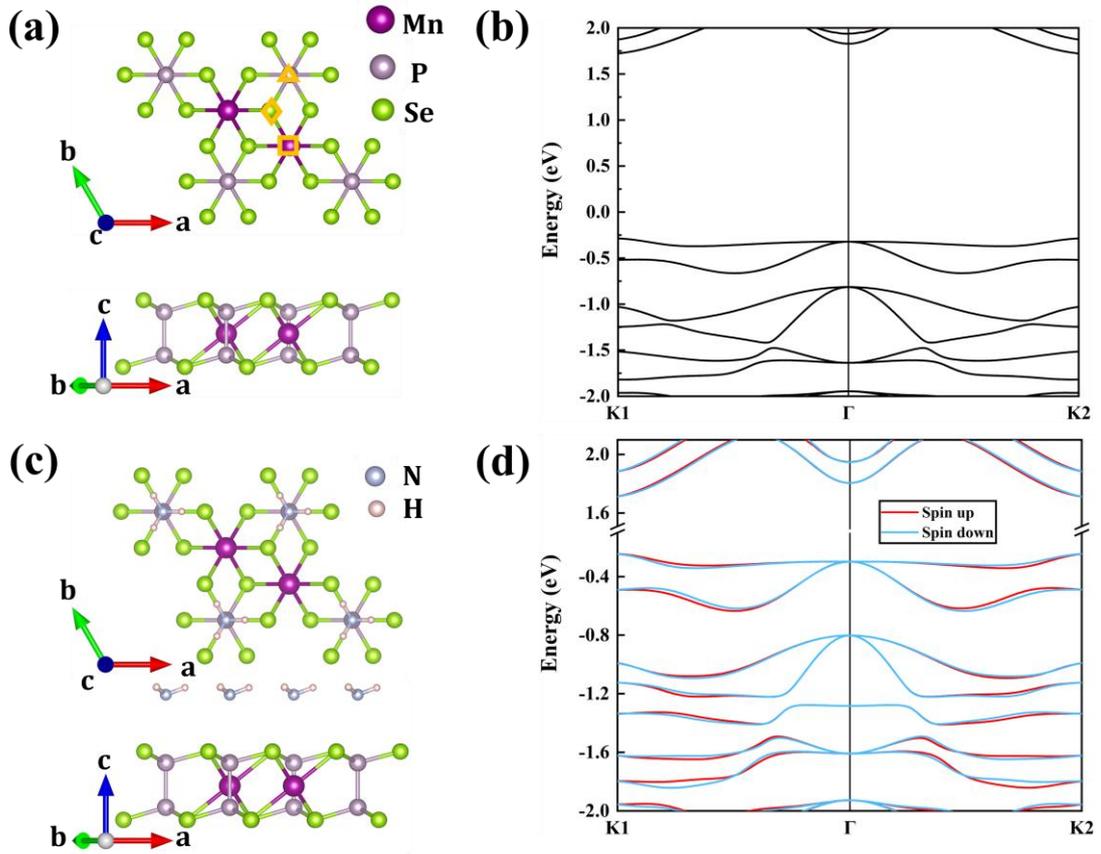

Fig. 5 (a) Top and side views of the initial monolayer $MnPSe_3$ structure. (b) Corresponding band structure without SOC. (c) Most energetically favorable configuration for $NH_3$ molecule adsorption on the monolayer $MnPSe_3$ surface at 100% coverage. (d) Corresponding spin-splitting band structure without SOC. The square, triangle and rhombus symbols denote the three considered adsorption sites of Mn, P and Se. The K path used to calculate the band structure in (b) and (d) is shown in Fig. 2(d).

## V. CONCLUSION

In summary, we systematically classifying all symmetry operations across the 80 layer groups to derive 63 distinct antiferromagnetic SPGs. Analysis of these groups identified 15 SPGs where the symmetry operations protecting spin degeneracy can be broken via surface adsorption—while preserving specific rotational or mirror symmetries connecting opposite-spin sublattices. This preservation is essential for realizing alternating spin polarization in antiferromagnetic materials. Demonstrating this principle, we achieved alternating spin-splitting in the originally spin-degenerate band structures of monolayer $MnPSe_3$ and monolayer $VPS_3$. This was accomplished by adsorbing

$NH_3$ molecules and O atoms, respectively. The feasibility and thermodynamic stability of these adsorption configurations were rigorously assessed through adsorption energy calculations across varying coverages and chemical potential phase diagrams. Both systems exhibit robust stability post-adsorption. Beyond providing concrete candidate materials for experimental realization of two-dimensional alternating magnets, this work introduces a vital theoretical framework. It elucidates the fundamental principle for generating alternating magnetism from two-dimensional antiferromagnets and presents a unified approach for significantly expanding the pool of potential materials exhibiting tunable magnetic order.

ACKNOWLEDGMENTS

This work was supported by the Guangdong Basic and Applied Basic Research Foundation (No. 2023A1515012289), and National Natural Science Foundation of China (Grant No. 12474229). This work is partially supported by High Performance Computing Platform of South China University of Technology.

(1) Šmejkal, L.; Sinova, J.; Jungwirth, T. Beyond Conventional Ferromagnetism and Antiferromagnetism: A Phase with Nonrelativistic Spin and Crystal Rotation Symmetry. *Phys. Rev. X* **2022**, *12* (3), 031042.

(2) Šmejkal, L.; Sinova, J.; Jungwirth, T. Emerging Research Landscape of Altermagnetism. *Phys. Rev. X* **2022**, *12* (4), 040501.

(3) Feng, Z.; Zhou, X.; Šmejkal, L.; Wu, L.; Zhu, Z.; Guo, H.; González-Hernández, R.; Wang, X.; Yan, H.; Qin, P.; Zhang, X.; Wu, H.; Chen, H.; Meng, Z.; Liu, L.; Xia, Z.; Sinova, J.; Jungwirth, T.; Liu, Z. An Anomalous Hall Effect in Altermagnetic Ruthenium Dioxide. *Nature Electronics* **2022**, *5* (11), 735–743.

(4) Zhang, S.-B.; Hu, L.-H.; Neupert, T. Finite-Momentum Cooper Pairing in Proximitized Altermagnets. *Nature Communications* **2024**, *15* (1), 1801.

(5) Ding, J.; Jiang, Z.; Chen, X.; Tao, Z.; Liu, Z.; Li, T.; Liu, J.; Sun, J.; Cheng, J.; Liu, J.; Yang, Y.; Zhang, R.; Deng, L.; Jing, W.; Huang, Y.; Shi, Y.; Ye, M.; Qiao, S.; Wang, Y.; Guo, Y.; Feng, D.; Shen, D. Large Band Splitting in $g$-Wave Altermagnet CrSb. *Phys. Rev. Lett.* **2024**, *133* (20), 206401.

(6) Ma, H.-Y.; Hu, M.; Li, N.; Liu, J.; Yao, W.; Jia, J.-F.; Liu, J. Multifunctional Antiferromagnetic Materials with Giant Piezomagnetism and Noncollinear Spin Current. *Nature Communications* **2021**, *12* (1), 2846.

(7) Sødequist, J.; Olsen, T. Two-Dimensional Altermagnets from High Throughput Computational Screening: Symmetry Requirements, Chiral Magnons, and Spin-Orbit Effects. *Applied Physics Letters* **2024**, *124* (18), 182409.

(8) Mazin, I.; González-Hernández, R.; Šmejkal, L. Induced Monolayer Altermagnetism in MnP(S, Se)$_3$ and FeSe. *arXiv:2309.02355,* **2023**.

(9) Ouassou, J. A.; Brataas, A.; Linder, J. Dc Josephson Effect in Altermagnets. *Phys. Rev. Lett.* **2023**, *131* (7), 076003.

(10) Bai, H.; Zhang, Y. C.; Zhou, Y. J.; Chen, P.; Wan, C. H.; Han, L.; Zhu, W. X.; Liang, S. X.; Su, Y. C.; Han, X. F.; Pan, F.; Song, C. Efficient Spin-to-Charge Conversion via Altermagnetic Spin Splitting Effect in Antiferromagnet $RuO_2$. *Phys. Rev. Lett.* **2023**, *130* (21), 216701.

(11) Gonzalez Betancourt, R. D.; Zubáč, J.; Gonzalez-Hernandez, R.; Geishendorf, K.; Šobáň, Z.; Springholz, G.; Olejník, K.; Šmejkal, L.; Sinova, J.; Jungwirth, T.; Goennenwein, S. T. B.; Thomas,

A.; Reichlová, H.; Železný, J.; Kriegner, D. Spontaneous Anomalous Hall Effect Arising from an Unconventional Compensated Magnetic Phase in a Semiconductor. *Phys. Rev. Lett.* **2023**, *130* (3), 036702.

(12) Lee, S.; Lee, S.; Jung, S.; Jung, J.; Kim, D.; Lee, Y.; Seok, B.; Kim, J.; Park, B. G.; Šmejkal, L.; Kang, C.-J.; Kim, C. Broken Kramers Degeneracy in Altermagnetic MnTe. *Phys. Rev. Lett.* **2024**, *132* (3), 036702.

(13) Yuan, L.-D.; Wang, Z.; Luo, J.-W.; Rashba, E. I.; Zunger, A. Giant Momentum-Dependent Spin Splitting in Centrosymmetric Low-$Z$ Antiferromagnets. *Phys. Rev. B* **2020**, *102* (1), 014422.

(14) Šmejkal, L.; MacDonald, A. H.; Sinova, J.; Nakatsuji, S.; Jungwirth, T. Anomalous Hall Antiferromagnets. *Nature Reviews Materials* **2022**, *7* (6), 482–496.

(15) Samanta, K.; Ležaić, M.; Merte, M.; Freimuth, F.; Blügel, S.; Mokrousov, Y. Crystal Hall and Crystal Magneto-Optical Effect in Thin Films of $SrRuO_3$. *Journal of Applied Physics* **2020**, *127* (21), 213904.

(16) Šmejkal, L.; Hellenes, A. B.; González-Hernández, R.; Sinova, J.; Jungwirth, T. Giant and Tunneling Magnetoresistance in Unconventional Collinear Antiferromagnets with Nonrelativistic Spin-Momentum Coupling. *Phys. Rev. X* **2022**, *12* (1), 011028.

(17) Wang, D.; Wang, H.; Liu, L.; Zhang, J.; Zhang, H. Electric-Field-Induced Switchable Two-Dimensional Altermagnets. *Nano Lett.* **2025**, *25* (1), 498–503.

(18) Chakraborty, A.; González Hernández, R.; Šmejkal, L.; Sinova, J. Strain-Induced Phase Transition from Antiferromagnet to Altermagnet. *Phys. Rev. B* **2024**, *109* (14), 144421.

(19) Zeng, S.; Zhao, Y.-J. Bilayer Stacking $A$-Type Altermagnet: A General Approach to Generating Two-Dimensional Altermagnetism. *Phys. Rev. B* **2024**, *110* (17), 174410.

(20) Liu, P.; Li, J.; Han, J.; Wan, X.; Liu, Q. Spin-Group Symmetry in Magnetic Materials with Negligible Spin-Orbit Coupling. *Phys. Rev. X* **2022**, *12* (2), 021016.

(21) Zeng, S.; Zhao, Y.-J. Description of Two-Dimensional Altermagnetism: Categorization Using Spin Group Theory. *Phys. Rev. B* **2024**, *110* (5), 054406.

(22) Bai, L.; Zhang, R.; Feng, W.; Yao, Y. Anomalous Hall Effect in Type IV 2D Collinear Magnets. arXiv:2504.08197.

(23) Kopsky, V.; Litvin, D. International Tables for Crystallography. *Vol. E, Subperiodic Groups,* **2002**.

(24) Litvin, D. B.; Opechowski, W. Spin Groups. *Physica* **1974**, *76* (3), 538–554.

(25) Liu, C.; Li, Z.; Hu, J.; Duan, H.; Wang, C.; Cai, L.; Feng, S.; Wang, Y.; Liu, R.; Hou, D.; Liu, C.; Zhang, R.; Zhu, L.; Niu, Y.; Zakharov, A. A.; Sheng, Z.; Yan, W. Probing the Néel-Type Antiferromagnetic Order and Coherent Magnon–Exciton Coupling in Van Der Waals $VPS_3$. *Advanced Materials* **2023**, *35* (30), 2300247.

(26) Wiedenmann, A.; Rossat-Mignod, J.; Louisy, A.; Brec, R.; Rouxel, J. Neutron Diffraction Study of the Layered Compounds $MnPSe_3$ and $FePSe_3$. *Solid State Communications* **1981**, *40* (12), 1067–1072.

(27) Klingen, W.; Eulenberger, G.; Hahn, H. Über Hexachalkogeno-Hypodiphosphate Vom Typ $M_2P_2X_6$. *Naturwissenschaften* **1970**, *57* (2), 88–88.

(28) Liu, D.; Zeng, S.; Liao, J.-H.; Zhao, Y.-J. Controllable Antiferromagnetic-Ferromagnetic Phase Transition in Monolayer $MnPSe_3$ via Atomic Adsorption of Li, O, and F. *arXiv:2503.11442,* **2025.**

(29) Li, X.; Wu, X.; Yang, J. Half-Metallicity in $MnPSe_3$ Exfoliated Nanosheet with Carrier Doping. *J. Am. Chem. Soc.* **2014**, *136* (31), 11065–11069.